\documentstyle[twoside,fleqn,espcrc2]{article}
\input{psfig}
\newcommand{\AmS}{{\protect\the\textfont2
  A\kern-.1667em\lower.5ex\hbox{M}\kern-.125emS}}

\begin{document}
\title{The diffractive structure function at the Tevatron: CDF results}
%\author{P. de Groot\address{Mathematics and Computer Science Division, 
%        Elsevier Science Publishers B.V., \\ 
%        P.O. Box 103, 1000 AC Amsterdam, The Netherlands}%
%        \thanks{Footnotes should appear on the first page only to
%                indicate your present address (if different from your
%                normal address), research grant, sponsoring agency, etc.
%                These are obtained with the {\tt\ttbs thanks} command.}
%        and 
%        X.-Y. Wang\address{Economics Department, University of Winchester, \\
%        2 Finch Road, Winchester, Hampshire P3L T19, United Kingdom}}
\author{K. Goulianos\address{The Rockefeller University,\\
1230 York Avenue, New York, NY 10021, USA\\
(dino@physics.rockefeller.edu)}%
\thanks{Representing the CDF Collaboration at 
``Diffraction 2000, Cetraro, Italy, 2-7 September 2000".}}      

\begin{abstract}
Results on hard diffraction from CDF are reviewed 
with emphasis on the 
determination of the diffractive structure function of the (anti)proton 
from single diffractive and double Pomeron exchange dijet production. 
Comparison of the diffractive dijet results with predictions based on 
diffractive deep inelastic scattering at HERA shows a breakdown of 
conventional QCD factorization.
A similar breakdown is observed within the CDF results in comparing 
the structure functions obtained from single diffraction and double 
Pomeron exchange. 
\end{abstract}

\maketitle

\section{INTRODUCTION}
\begin{figure}[h,t]
\vglue -0.5in
{\hspace*{-0.5in}\psfig{figure=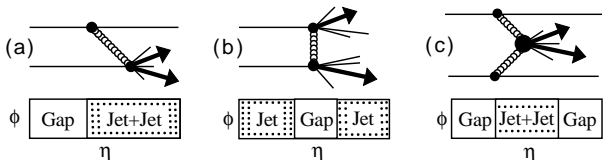,width=4.2in}}
\vglue -5in
\caption{Dijet production diagrams and event topologies for 
(a) single diffraction (b) double diffraction
and (c) double Pomeron exchange.}
\label{topology}
\end{figure}

Hadronic diffraction~\cite{credits} is believed to be mediated by the exchange 
of the Pomeron~\cite{Regge}. 
In the framework of QCD, the Pomeron consists of (anti)quarks and gluons 
in a color-singlet state with vacuum quantum numbers.
An interesting question is whether the Pomeron, although virtual, 
has a {\em unique} partonic structure obeying QCD factorization,
as is the case for real hadrons. Such a structure could be probed
in diffractive processes which incorporate a hard scattering.
Figure~\ref{topology} shows the event topology for dijet production in 
single diffraction (SD), double diffraction (DD),
and double Pomeron exchange (DPE).
\vglue -0.3in

\begin{figure}[h]
\vglue -1.4in
\hspace*{-0.62in}\psfig{figure=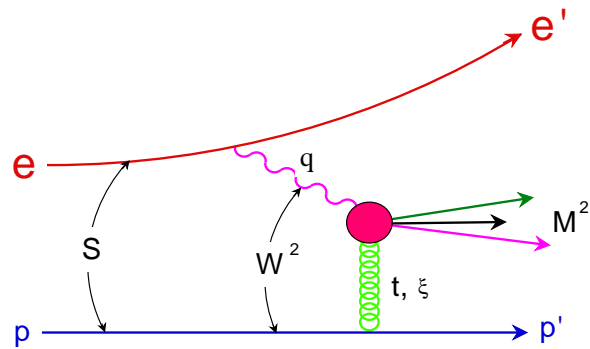,width=4.5in}
\vglue -3.5in
\caption{Schematic diagram of a diffractive
DIS collision involving a virtual photon,
emitted by an electron, and a virtual Pomeron,
emitted by a proton.}
\label{HERA}
\vglue -0.1in
\end{figure}

Since there is no color exchanged
between the colorless Pomeron and the parent nucleon,
a rapidity gap (region devoid of particles)
emerges as a characteristic signature of Pomeron exchange, as shown
schematically in the $\eta-\phi$ plots of Fig.~\ref{topology}. 
Such gaps can be used
to tag diffractive production. Another way of tagging diffraction 
is provided by the recoil $\bar p$
or $p$ in SD or in DPE. The CDF Collaboration has reported results for 
hard diffraction in $\bar pp$ collisions at $\sqrt{s}=1800$ GeV 
obtained by using both tagging 
techniques~\cite{CDF_W,CDF_JJG,CDF_B,CDF_JJ,CDF_DPE}.\\
%%%%%%%%%%%%%%%
% for CDF note
\vspace*{-5.25in}
\begin{flushright}
%\fbox{CDF/PUB/CDF/PUBLIC/5464}\\
\fbox{RU 00/E-17}\\
\vglue 0.3in
%\vglue 0.25cm
\today\\
\end{flushright}
\vspace*{4.25in}
%%%%%%%%%%%%%%%
At HERA, where $\sim 28$ GeV electrons are brought into collision with
$\sim 800$ GeV protons ($\sqrt{s}\approx 300$ GeV),
diffraction has been studied both in photoproduction and
in high $Q^2$ deep inelastic scattering (DIS). The H1 and ZEUS
Collaborations have measured the diffractive structure function of the
proton and its factorization properties~\cite{H1,ZEUS}. 
Figure~\ref{HERA} shows the 
kinematics of a DIS diffractive collision.\\

In analogy with the $F_2(Q^2,x)$ structure function,
the 4-variable diffractive structure function (DSF) of the proton,
$F_2^{D(4)}(Q^2,x,\xi,t)$, 
is defined through the cross section  equation
\begin{equation}
\frac{d^4\sigma}{dQ^2\,dx\, d\xi\, dt}=\frac{4\pi \alpha^2}{x\, Q^4}\cdot
f(y)\cdot F_2^{D(4)}(Q^2,x,\xi,t)
\label{F2D4}
\end{equation}
where $x$ and $y$ are the Bjorken DIS variables, 
$f(y)\equiv 2-2y+{y^2}/[{2(1+R)}]$, and $\xi$ is the fraction of the 
proton's momentum carried by the Pomeron.
The $t$-integrated DSF,
$F_2^{D(3)}$,
has been determined from the data by assuming $R=0$ in $f(y)$. 
The reported measured values for $F_2^{D(3)}$
correspond to cross sections given by 
\begin{equation}
\frac{d^3\sigma}{dQ^2 \,d\beta \,d\xi}\,\stackrel{|t|<1}{\equiv}\,
\frac{4\pi \alpha^2}{\beta Q^4}\cdot f(y)\cdot F_2^{D(3)}(Q^2,\beta,\xi)
\label{F2D3}
\end{equation}
where $\beta\equiv x/\xi$ represents the fraction of the momentum of 
the Pomeron carried by the interacting quark.

Diffractive quark densities are derived directly from $F_2^{D(3)}$. 
Using a QCD analysis, the H1 Collaboration has also 
derived~\cite{H1} diffractive
gluon densities from the $Q^2$ dependence (scaling violations)
of $F_2^{D(3)}$. In this paper, we present CDF results on 
single-diffractive to non-diffractive (ND) ratios for
$W$, dijet, $b$-quark and $J/\psi$ production measured by using rapidity gap 
tagging, as well for dijet production 
using leading $\bar p$ tagging, and compare them with expectations based 
on diffractive parton densities extracted from HERA measurements 
to test factorization.  In addition, 
factorization is tested within the CDF data by comparing dijet production 
in SD and in DPE.

\section{RESULTS USING RAPIDITY GAPS}
Using the rapidity gap method to tag diffraction, four hard SD processes 
have been studied by CDF: Diffractive $W$~\cite{CDF_W},
dijet~\cite{CDF_JJG}, $b$-quark~\cite{CDF_B} and $J/\psi$ production.

The components of the CDF detector relevant to forward rapidity gap 
tagging are~\cite{CDF} the
Beam-Beam Counters (BBC)
and the forward calorimeters (FCAL).
The BBC consist of square arrays of 16 scintillation
counters placed at  $\pm z$ positions of 6 m from the center of the
detector. The FCAL have a tower geometry with segmentation of 0.1
units in $\eta$ and $5^{\circ}$ 
in $\phi$. These detectors cover the $\eta$-range:

\begin{tabular}{lr}
BBC & $~3.2<|\eta|<5.9$\\
FCAL&$2.4<|\eta|<4.2$\\
\end{tabular}

\noindent Experimentally, a ``particle" is
defined as a hit in the BBC or a calorimeter tower with energy 
$E>1.5$ GeV. The tower energy threshold is imposed to reduce 
calorimeter noise and to enhance the correspondence between 
a single tower above threshold and a real particle. 
In a sample of $W$, dijet, $b-$quark or $J/\psi$ events, 
the SD signal is identified 
as an excess of events above a ND ``background" in the (0,0) bin
of the BBC versus FCAL multiplicity distribution. The ratio of SD to ND 
events is then evaluated, compared with Monte Carlo simulations, and 
conclusions are drawn about the Pomeron structure 
function and its factorization properties. Below, we briefly present  
the results obtained for each process studied by CDF 
and the conclusions drawn about the Pomeron structure.

\subsection{Diffractive $W$ production}
CDF made the first observation~\cite{CDF_W}  
of diffractive $W$s and measured the $W$ production 
rate using a sample of 8246 events with an isolated 
central $e^+$ or $e^-$ ($|\eta|<1.1$)  of $E_T>20$ GeV 
and missing transverse energy $\not\!\! E_T>20$ GeV. In searching for
diffractive events, CDF studied the correlations of 
the BBC multiplicity, $N_{BBC}$, 
with the sign of the electron-$\eta$, $\eta_e$,  or  
the sign of its charge, $C_e$. 
In a diffractive $W^{\pm}\rightarrow e^{\pm}\nu$ event produced in
a $\bar p$ collision with a Pomeron emitted by the proton,
a rapidity gap is expected at positive $\eta$ ($p$-direction),
while the lepton is boosted towards negative $\eta$ (angle-gap correlation).
Also, since the Pomeron is quark-flavor symmetric,
and since, from energy considerations, mainly valence quarks from the
$\bar p$ participate in producing the $W$,
approximately twice as many electrons as positrons
are expected (charge-gap correlation). 

\begin{figure}[htp]
\vglue -1cm
\psfig{figure=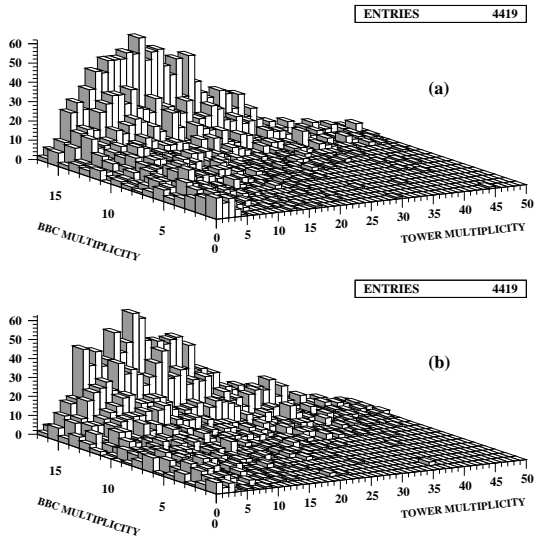,width=3.2in}
\vglue -1cm
\caption{Tower versus BBC multiplicity for $W$ events: 
(a) (angle$\otimes$charge)-correlated;
(b) (angle$\otimes$charge)-anticorrelated.}
\label{FIG_W_1}
\vglue -0.5cm
\end{figure}

Figure~\ref{FIG_W_1} shows the BBC versus tower multiplicity 
for two event samples characterized by the correlation between the 
pseudorapidity of the BBC whose multiplicity is plotted, $\eta_{BBC}$,
and the $\eta_e$ or $C_e$, as follows:  (a)  
{\em doubly-correlated} events, for which 
$\eta_e\cdot C_e>0$ and $\eta_e\cdot \eta_{BBC}<0$, and (b) 
{\em doubly-anticorrelated} events, for which
$\eta_e\cdot C_e>0$ and $\eta_e \cdot \eta_{BBC}>0$. 
Monte Carlo simulations show that 
diffractive $W$ events are expected to have low BBC or tower 
multiplicities (in the range 0-3), and that there should be about 4 times as 
many doubly-correlated than doubly-anticorrelated events. 
This diffractive signature is satisfied by the small number of events 
at low multiplicities in Fig.~\ref{FIG_W_1}. 

\begin{figure}[htp]
\vglue -1.5cm
{\hspace*{-0.3cm}\psfig{figure=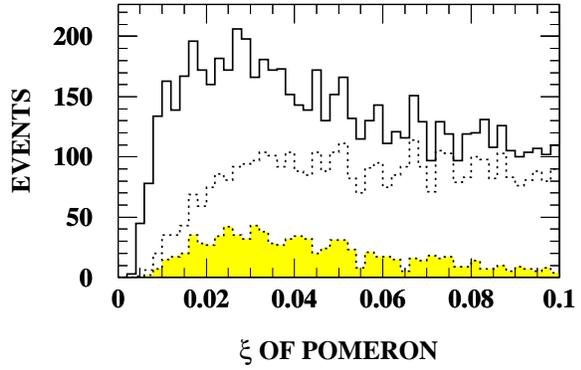,width=6.6in}
}
\vglue -11.5cm
\caption{
Monte Carlo Pomeron $\xi$ distributions for diffractive
$W$ events generated by POMPYT
using a hard-quark Pomeron structure: {\em (solid line)} 
all events; {\em (dotted line)} 
events with a central electron; {\em (shaded area)} events
with a central electron and 0, 1 or 2 hits in the 
(angle$\oplus$charge)-correlated BBC (corresponding to the signal).}
\label{FIG_W_3}
\vglue -0.5cm
\end{figure}

Correcting for acceptance,
the ratio of diffractive to non-diffractive $W$ production is:\\
\centerline{$R_W=[1.15\pm 0.51(stat)\pm 0.20(syst)]\%\;\,(\xi<0.1)$}\\
Figure~\ref{FIG_W_3} shows Pomeron-$\xi$ distributions of $W$
events generated by the POMPYT~\cite{POMPYT} Monte Carlo program followed 
by a detector simulation.

\subsection{Diffractive dijet production}
\begin{figure}[htp]
\vglue -3.5cm
\psfig{figure=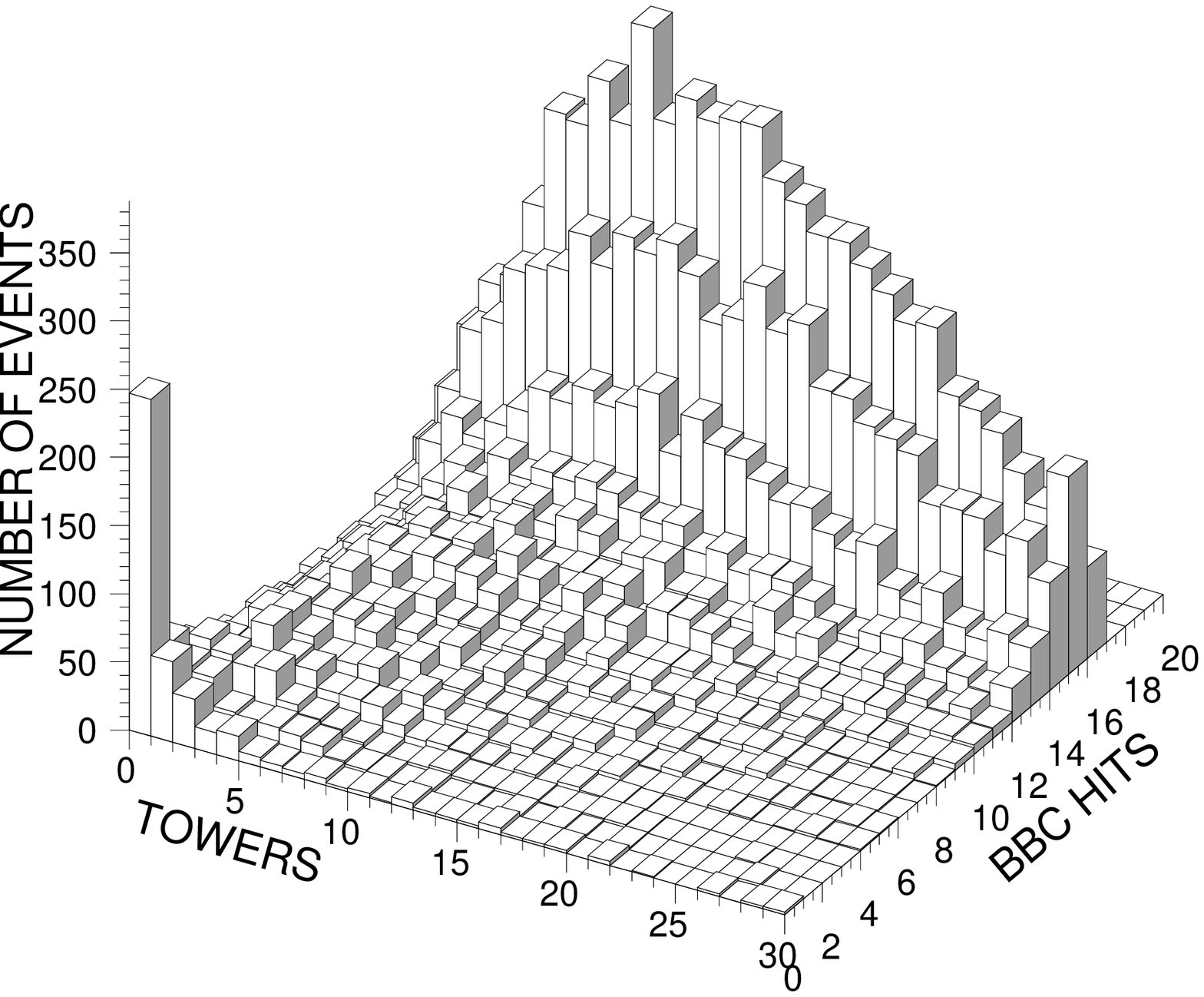,width=3.2in}
\vglue -1cm
\caption{FCAL tower versus BBC multiplicity on the opposite hemisphere of the 
dijet system for dijet events with both jets at
$\eta>1.8$ or $\eta<1.8$.}
\vglue -1cm
\label{FIG_GJJ_1}
\end{figure}

\begin{figure}[htp]
\vglue -3cm
{\hspace*{-0.3cm}\psfig{figure=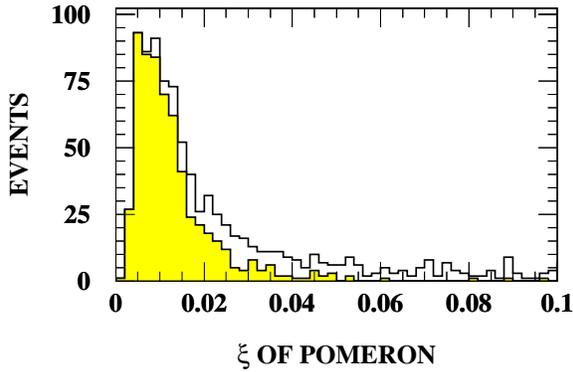,width=6.6in}}
\vglue -11cm
\caption{
Monte Carlo Pomeron $\xi$ distributions for diffractive
dijet events with jet $E_T>20$ GeV and $1.8<|\eta|<3.5$ generated by POMPYT
using a hard-gluon Pomeron structure.  The shaded area
represents the subset of Monte Carlo events
with zero BBC and forward calorimeter multiplicities, corresponding to the
data in the (0,0) bin of Fig.~\ref{FIG_GJJ_1}.}
\label{FIG_GJJ_2}
\vglue -2cm
\end{figure}

CDF searched for diffractive dijet production in a sample of 30352 dijet 
events with a single-vertex 
(to exclude events from multiple interactions), in which  
the two leading jets 
have $E_T>20$ GeV and are both at $\eta<1.8$ or $\eta>1.8$.
No requirement was imposed on the presence or kinematics of extra jets in 
an event. Figure~\ref{FIG_GJJ_1} shows the correlation of the BBC and forward 
($|\eta|>2.4$) calorimeter 
tower multiplicities in the $\eta$-region opposite the dijet system. The 
excess in the 0-0 bin is attributed to diffractive production. 
After subtracting the non-diffractive background and 
correcting for the single-vertex selection cut, for 
detector live-time acceptance  and for the 
rapidity gap acceptance $(0.70\pm 0.03)$, calculated using the POMPYT 
Monte Carlo program with Pomeron $\xi<0.1$, the 
``Gap-Jet-Jet" fraction (ratio of diffractive to non-diffractive dijet events)
was found to be 
\begin{center}
$R_{GJJ}=[0.75\pm 0.05(stat)\pm 0.09 (syst)]\%=(0.75\pm 0.10)\%$\\ 
($E_T^{jet}>20$ GeV, $|\eta|^{jet}>1.8$, $\eta_1\eta_2>0$, $\xi<0.1$)
\end{center}
Figure~\ref{FIG_GJJ_2} shows Pomeron-$\xi$ distributions of dijet 
events generated by a POMPYT Monte Carlo simulation for $\xi<0.1$ using 
a hard gluon Pomeron structure. The jets were required to have 
$E_T^{jet}>20$ GeV and be in the region $1.8<|\eta|<3.5$ with 
$\eta_1\cdot \eta_2 >0$.

\subsection{Diffractive $b$-quark production}
\begin{figure}
\vglue -1cm
\centerline{\psfig{figure=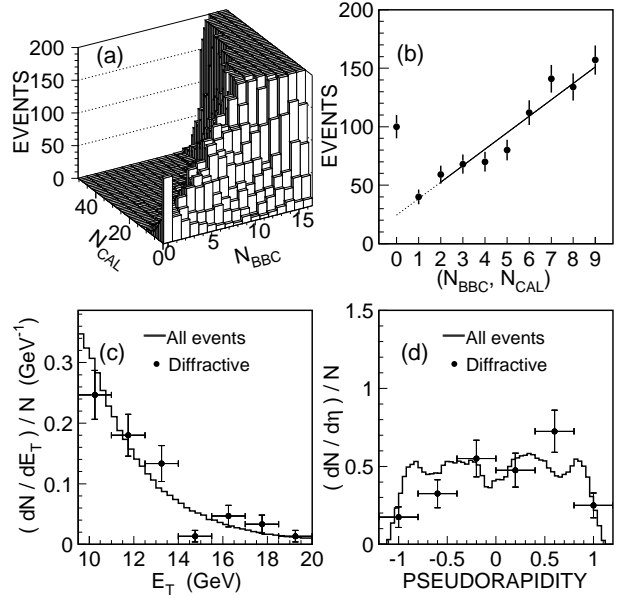,width=3.5in}}
\vglue -1cm
\caption{Diffractive $b$-quark production: 
(a) Forward calorimeter tower multiplicity, $N_{CAL}$, versus
beam-beam counter multiplicity, $N_{BBC}$;
(b) multiplicity distribution along the diagonal
with $N_{BBC}=N_{CAL}$ in the plot in (a);
(c) electron $E_T$; and (d) pseudorapidity for the
diffractive (points) and total (histogram) event samples
(diffractive events with a rapidity gap at positive $\eta$ are entered
with the sign of the electron $\eta$ changed).}
\vglue -0.5cm
\label{fig:b1}
\end{figure}

Diffractive $b$-quark production probes directly the gluon content of the 
Pomeron. In a sample of events collected in the 1994-95 run (80 pb$^{-1}$)
with a trigger requiring an electron of $E_T^e>7.5$ GeV within $|\eta|<1.1$,
CDF searched for diffractive events after imposing the software cuts 
of $E_T^e>9.5$ GeV (to avoid trigger bias) and $E_T^e<20$ GeV (to reject $W$ 
and $Z$ boson events. 
As in the previous studies, the diffractive signal was evaluated 
from the FCAL versus BBC multiplicity distribution, which is shown in 
Fig.~\ref{fig:b1}a. The excess of events in the (0,0) bin above a smooth 
extrapolation from nearby bins is attributed to diffractive events. 
The ND content of the (0,0) bin was evaluated from the distribution of events 
along the diagonal of Fig.~\ref{fig:b1}a with $N_{BBC}=N_{CAL}$, which is 
shown in 
Fig.~\ref{fig:b1}b. Figures \ref{fig:b1}c and \ref{fig:b1}d show the electron 
$E_T$ and $\eta$ distributions, respectively, for the diffractive and total 
event samples.

After extracting the $b$-quark content from the event sample, separately for 
the SD and ND events, and correcting for various backgrounds~\cite{CDF_B},
the ratio of SD to ND $b$ production was found to be 
$$R^{\rm gap}_{\bar bb}=[0.23\pm 0.07({stat})\pm 0.05({syst})]\%$$
The rapidity gap acceptance for events generated with a flat Pomeron structure,
which is approximately the structure obtained from the 
HERA measurements~\cite{H1,ZEUS},
and a gluon fraction of $0.7\pm 0.2$,
as reported in~\cite{CDF_JJG}, is found to be $0.37\pm 0.02$.
Dividing $R^{\rm gap}_{\bar bb}$ by this value yields a
diffractive to total production ratio of 
$$R_{\bar bb}=[0.62\pm 0.19(stat)\pm 0.16(syst)]\%$$
$$(9.5<E_T^E<20\;{\rm GeV}\;\;\;|\eta^e|<1.1\;\;\;\xi<0.1)$$
\begin{figure}[t]
\vglue -1.5cm
\centerline{\psfig{figure=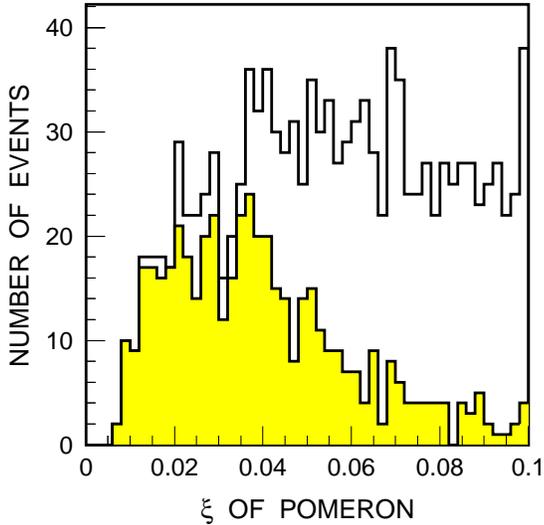,width=3.5in}}
\vglue -1cm
\caption{Monte Carlo distribution of the Pomeron beam momentum fraction, $\xi$,
for diffractive $b$-quark production events with an electron
of $9.5<p_T^e<20$ GeV/$c$ within $|\eta|<1.1$,
generated using a flat Pomeron structure
of gluon to quark ratio $0.7\div 0.3$.
The shaded area is the distributions for
events satisfying the rapidity gap requirements.}
\vglue -0.5cm
\label{fig:b2}
\end{figure}
Fig.~\ref{fig:b2} shows Pomeron-$\xi$ distributions of $b$
events generated by the POMPYT Monte Carlo program followed 
by a detector simulation.

\subsection{Diffractive $J/\psi$ production}
\begin{figure}[h]
\vglue -0.5cm
\centerline{\psfig{figure=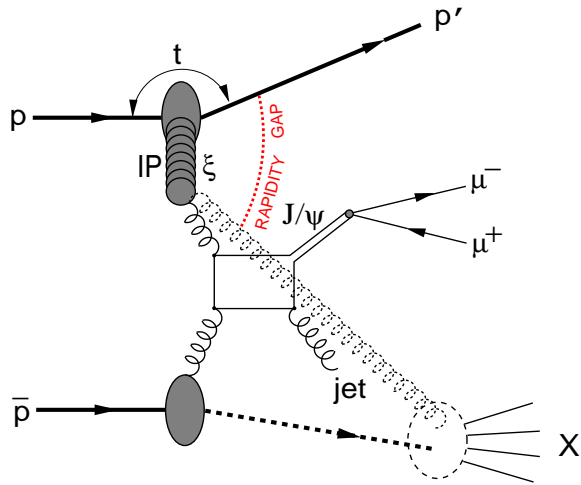,width=3in}}
\vglue -1cm
\caption{Schematic diagram of diffractive $J/\psi$ production.
Note the jet in the final state, which is expected to balance the
$J/\psi$ momentum in diffractive events.}
%\vglue -1cm
\label{fig:jpsi1}
\end{figure}

As in the case of diffractive $b$-quark production,
diffractive $J/\psi$ production probes {\em directly}
the gluon content of the Pomeron (see Fig.~\ref{fig:jpsi1}).
The $J/\psi$ study is interesting because, since muons from
$J/\psi\rightarrow \mu^+\mu^-$ can be detected in CDF down to $p_T\sim 2$ GeV,
the SD to ND
fraction can be measured at $p_T$ values about five times lower than those in
the $b$-quark case.
Thus, from the combined $J/\psi$ and $b$-quark
measured fractions one may determine the $p_T$ dependence
of the gluon fraction of the Pomeron.
Moreover,  the leading (anti)proton momentum fraction of the
parton participating in $J/\psi$ production, $x_{bj}$,
can be determined from the measured vector momenta of the $J/\psi$
and the accompanying jet.
This allows one to determine
the fraction of SD to ND production as a function of $x_{bj}$.
\begin{figure}[t]
\vglue -1cm
\centerline{\psfig{figure=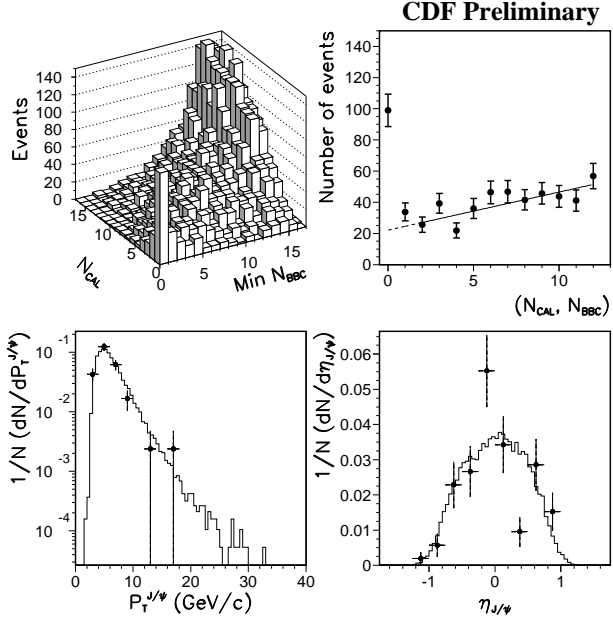,width=3.5in}}
\vglue -1cm
\caption{Diffractive $J/\psi$ production: 
(a) Forward calorimeter tower multiplicity, $N_{CAL}$, versus
beam-beam counter multiplicity for the $\eta$ side of the event 
with the minimum BBC  multiplicity, Min $N_{BBC}$;
(b) multiplicity distribution along the diagonal
with $N_{BBC}=N_{CAL}$;
(c) $p_T$ of $J/\psi$, and (d) pseudorapidity for the
diffractive (points) and total (histogram) event samples
(diffractive events with a rapidity gap at positive $\eta$ are entered
with the sign of the electron $\eta$ changed).}
\vglue -1cm
\label{fig:jpsi2}
\end{figure}
Figure~\ref{fig:jpsi2} shows the same distributions as those in 
Fig.~\ref{fig:b1} but for $J/\psi$ rather than $b$-quark events.
The $J/\psi$ is identified by its decay into two muons, 
$J/\psi\rightarrow \mu^+\mu^-$.
The events selected have muons of $p_T^{\mu}>2$ GeV within $|\eta|<1.1$. 
The SD to ND fraction for events with a forward rapidity gap is found to be 
$$R^{\rm gap}_{J/\psi}=[0.36\pm 0.06({\rm stat})]\%$$
This value is  
is about 1.5 times larger than the measured SD to ND $b$-quark 
fraction, $R^{\rm gap}_{\bar bb}$. Under the reasonable 
assumption that the gap acceptances are similar for the two cases, 
the larger $R$ value measured in $J/\psi$ production must be attributed to 
the lower $p_T$ of the hard scattering. 
\subsection{Gluon fraction and factorization}
\begin{figure}[t]
\vglue -1cm
\centerline{\psfig{figure=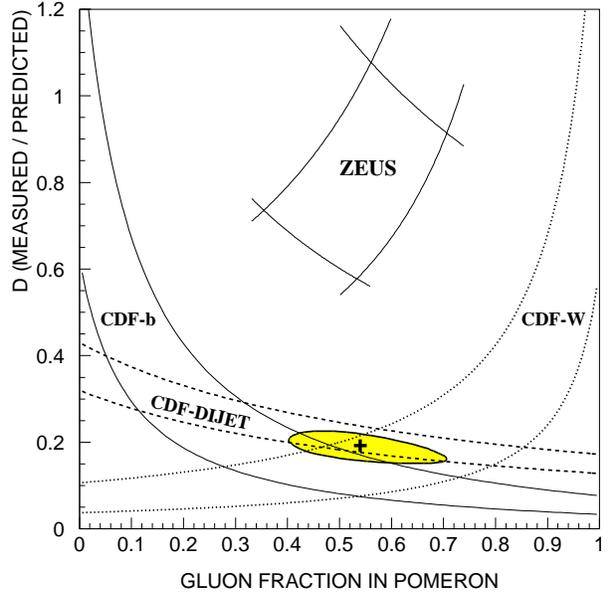,width=3.5in}}
\vglue -1cm
\caption{The ratio, $D$, of measured to predicted diffractive
rates as a function of the gluon content of the Pomeron.
The predictions are from POMPYT using the standard Pomeron flux
and a hard Pomeron structure.
The CDF-$W$ curves were calculated
assuming a three-flavor quark structure for the Pomeron.
The black cross and shaded ellipse
are the best fit and $1\sigma$ contour of a least square two-parameter fit
to the three CDF results.}
\vglue -1cm
\label{fig:f_g}
\end{figure} 
By combining the diffractive $W$, dijet and $b$-quark results, CDF extracted 
the gluon fraction of the Pomeron, $f_g$. 
Assuming the standard Pomeron flux in the POMPYT Monte Carlo program, 
the measured $W$, dijet and $b$-quark SD to ND ratios  
trace  curves in the plane of $D$ versus $f_g$, where $D$ is 
the ratio of measured to POMPYT-predicted rates. 
Figure~\ref{fig:f_g}
shows the $\pm1\sigma$ curves corresponding to the results. 
From the oval-shaped overlap of the $W$, dijet and $b$-quark
curves,  CDF obtained 
$f_g=0.54^{+0.16}_{-0.14}$. This result, which is independent of the 
Pomeron flux normalization, agrees with the result obtained by 
ZEUS~\cite{ZEUS}
from DIS and dijet photoproduction (diamond-shaped area shown in the figure).
For the $D$-fraction, CDF obtained the value 
$D=0.19\pm 0.04$. 
The decrease of the 
$D$-fraction from HERA to the Tevatron
represents a breakdown of factorization. 

The observed factorization breakdown can be further characterized 
by measuring the {\em shape} of the diffractive structure
function at the Tevatron and comparing it with predictions based on 
the HERA measurements. This has been done by CDF in studies of 
dijet production in SD and DPE using leading particle tagging,
as described in the next section.

\section{RESULTS USING ROMAN POTS}
Using a Roman Pot Spectrometer (RPS) to trigger on and measure the 
momentum of leading antiprotons, CDF collected about 3,000 K events
at $\sqrt s=1800$ GeV with no other requirement on the trigger.
After requiring  
a single vertex within $|z_{vtx}|<60$ cm, 
to reject events due to more than one interactions, 
and applying quality cuts on the RPS tracks, there remained 1,639 K events 
in the region $0.035<\xi_{\bar p}<0.095$ and $|t|<1$~GeV$^2$.
This {\em inclusive diffractive} event sample contains 30,639 events 
with two jets of $E_T^{jet}>7$ GeV. A non-diffractive 
event sample was also collected
by requiring only a BBC coincidence. This {\em Minimum Bias} (MB) sample 
consists of 299,959 events with $|z_{vtx}|<60$ cm and contains 32,629
dijet events of $E_T^{jet}>7$ GeV.
The above event samples were used 
for measuring the diffractive structure function 
of the antiproton in SD events, as well as that of the proton in DPE.
Diffractive factorization was tested by comparing these structure function 
to each other and with predictions based on HERA measurements,
as described below.
\subsection{Single-diffractive dijets}
The diffractive dijet sample was used to measure the diffractive structure 
function (DSF) of the antiproton~\cite{CDF_JJ}. The procedure used, designed to 
avoid Monte Carlo simulations, is based on measuring the 
ratio $R(x)$ of SD to ND cross sections as a function of the Bjorken $x$ of the 
parton in the $\bar p$ participating in the hard scattering. In LO QCD, 
this ratio is proportional to the corresponding structure functions. 
The DSF is then obtained by multiplying the measured $R(x)$ by the known 
ND structure function.
The absolute normalization of the SD dijet sample is obtained
by scaling the event rate to that of the inclusive diffractive sample
and using for the latter the previously measured inclusive cross
section~\cite{CDFEDT}.
The normalization of the ND dijet sample is obtained from the measured
$51.2\pm 1.7$ mb cross section of the BBC trigger.

The variable $x$ is evaluated from the jet $E_T$ and $\eta$, as follows:
$$x=\frac{1}{\sqrt{s}}\sum_{i=1}^nE_T^ie^{-\eta^i}$$
The sum is carried
over the two leading jets plus the next highest
$E_T$ jet, if there is one with $E_T>5$ GeV.

\begin{figure}[h]
\vglue -1.5cm
\centerline{\psfig{figure=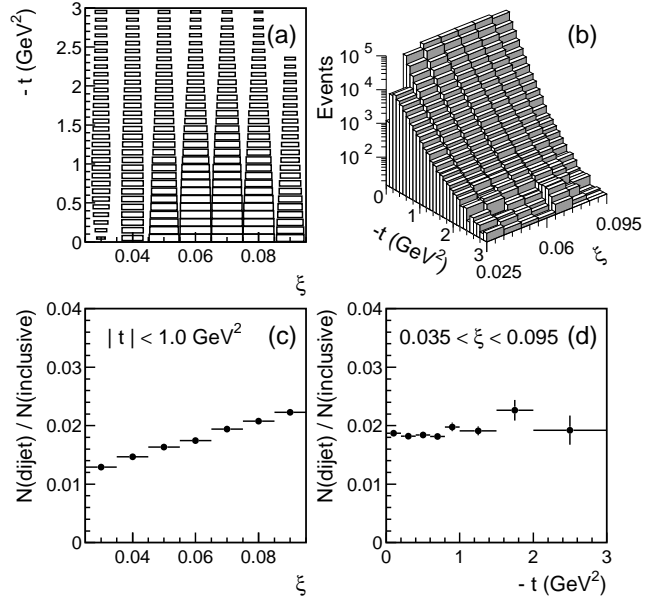,width=3.5in}}
\vglue -1cm
\caption{Distributions versus $\xi$ and $t$:
(a) Roman pot acceptance; (b) inclusive diffractive event sample;
(c) ratio of dijet to inclusive diffractive events versus $\xi$ and (d)
versus $t$.}
\label{fig:sd1}
\vglue -0.7cm
\end{figure}
The structure function relevant to dijet production is a color-weighted 
combination of quark and gluon components
$$F_{jj}(x)=x\left\{g(x)+\frac49\sum_i [(q_i(x)+{\bar q}_i(x)]\right\}$$
\noindent where $g(x)$ and $q(x)$ are gluon and quark parton densities,
respectively. For comparisons with predictions based on HERA results,
in which the DSF is usually presented in terms of the variable $\beta$ 
instead of $x$, the DSF obtained from the equation 
$F^D_{jj}(x,\xi)=R(x,\xi)\times F^{ND}_{jj}(x)$ may be transformed to 
${F}^D_{jj}(\beta,\xi)$ by a change of variables ($x=\beta \xi$).

\begin{figure}[t]
\vglue -0.7cm
\centerline{\psfig{figure=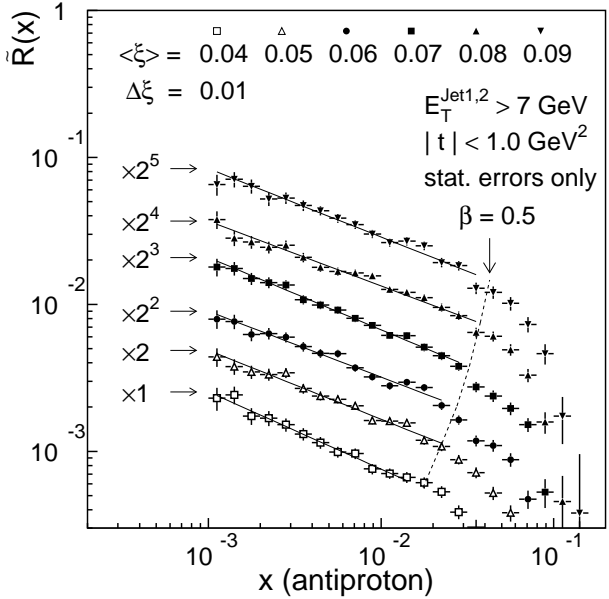,width=3.5in}}
\vglue -1cm
\caption{Ratio of diffractive to non-diffractive dijet event rates
as a function of
$x$ (momentum fraction of parton in $\bar p$).
The solid lines are fits to the form $\tilde{R}(x)=R_{\circ}(x/0.0065)^{-r}$
for $\beta<0.5$.}
\label{fig:sd3}
\vglue -0.7cm
\end{figure}

\begin{figure}[ht]
\vspace{-0.7cm}
\centerline{\psfig{figure=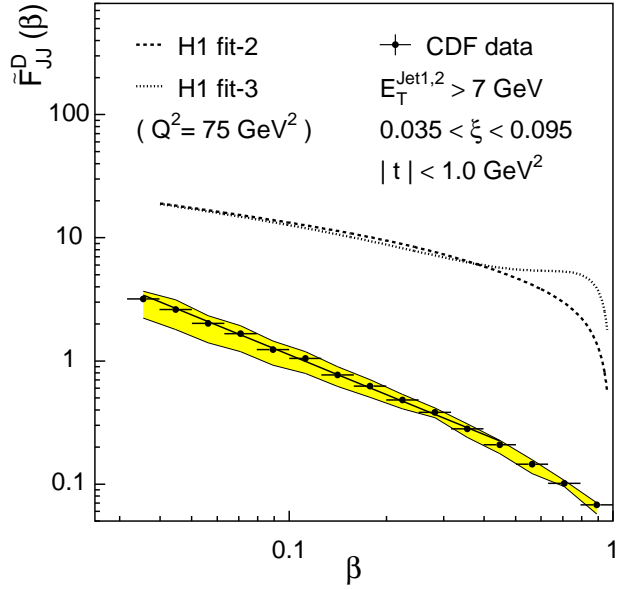,width=3.5in}}
\vglue -1cm
\caption{Data $\beta$ distribution (points) compared with expectations from
the parton densities of the proton extracted from diffractive deep
inelastic scattering by the H1 Collaboration at HERA. 
%The systematic uncertainty in the normalization of the data is $\pm 25\%$.
}
\label{fig:sd4}
\vglue -0.7cm
\end{figure}

Results of the diffractive dijet analysis are presented in 
Figs.~\ref{fig:sd1}, \ref{fig:sd3} and \ref{fig:sd4}. The highlights are:
(i) In Fig.~\ref{fig:sd1}d, the ratio of SD dijet to inclusive 
production rates is seen to be independent of $t$;
(ii) In Fig.~\ref{fig:sd3}, the ratio of SD to ND dijet 
production rates exhibits a power law 
dependence for $\beta<0.5$, increasing with 
decreasing $x$ as $x^{-0.46}$;
(iii) In Fig.~\ref{fig:sd4}, the diffractive structure function measured by CDF
differs both in shape and normalization from expectations based on 
diffractive parton densities extracted by the H1 Collaboration from 
diffractive DIS measurements. The normalization discrepancy, 
which is of ${\cal{O}}(0.1)$,
confirms the breakdown of factorization observed in the comparison of the 
rapidity gap results with expectations from HERA measurements.

\subsection{Double-Pomeron dijets}
\begin{figure}
\vglue -1cm
\centerline{
\hspace{0.3cm}\psfig{figure=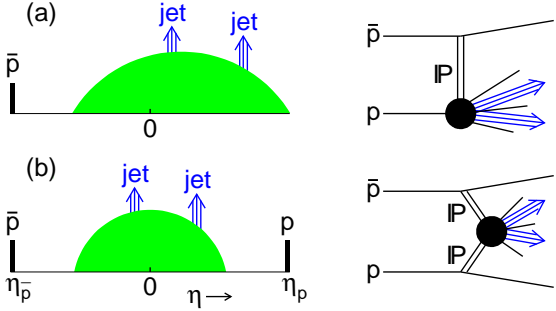,width=3.25in}
}
\vglue -1.8in
\caption{Illustration of event topologies in pseudorapidity, $\eta$,
and associated Pomeron exchange
diagrams for dijet production in (a) single diffraction and (b) double
Pomeron exchange. The shaded areas on the left side represent
particles not associated with the jets (underlying event).}
\label{fig:dpe1}
\vglue -2cm
\end{figure}

Using the RPS single-diffractive dijet event sample, 
CDF searched for and discovered~\cite{CDF_DPE} 
events with a rapidity gap on the proton side,
presumed to be due to the double Pomeron exchange 
process illustrated in Fig.~\ref{fig:dpe1}b. 

\begin{figure}[htp]
\vglue -0.2cm
\centerline{\psfig{figure=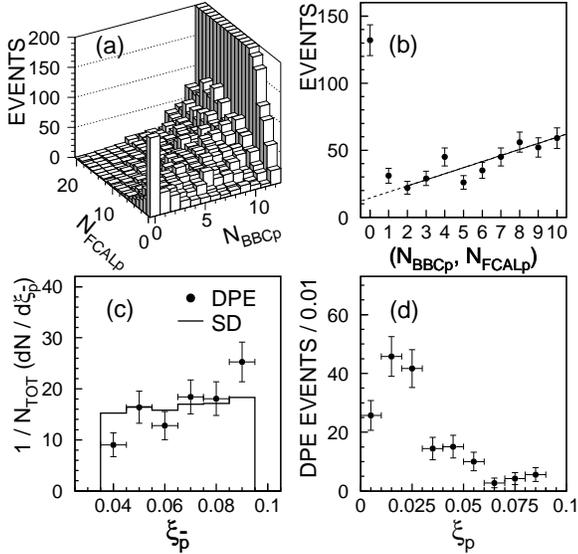,width=3.1in}}
\vglue -1cm
\caption{(a) Beam-Beam Counter hit multiplicity on the proton side,
$N_{BBCp}$, versus forward calorimeter tower multiplicity, $N_{FCALp}$:
the peak in the (0,0) bin contains the DPE signal;
(b) multiplicity distribution along the diagonal bins in (a) with
$N_{BBCp}=N_{FCALp}$; (c) $\xi_{\bar{p}}$ measured by
the RPS for SD events (histogram) and for the ``DPE" events of the (0,0)
bin in (a);
(d) $\xi_p$ of the DPE events. In (c) and (d) the data are
corrected for RPS acceptance on an event-by-event basis.}
\label{fig:dpe2}
\vglue -0.7cm
\end{figure}

The DPE signal appears as an 
enhancement in the (0,0) bin of the $N_{FCALp}$ versus $N_{BBCp}$ distribution
of Fig.~\ref{fig:dpe2}a, and in the one-dimensional ``diagonal" distribution
of Fig.~\ref{fig:dpe2}b.  The $\xi_p$ distribution
for the events of the (0,0) bin of Fig. \ref{fig:dpe2}a 
is shown in Fig.~\ref{fig:dpe2}d. 
The $\xi_p$ values were determined from the
calorimeter and BBC information~\cite{CDF_DPE}.

\newpage

In events with a leading antiproton, or equivalently with a rapidity gap 
on the $\bar p$ side, 
the ratio of the DPE to SD dijet production cross sections at the same 
$x_p$ for fixed $\xi_p$, $R^{DPE}_{SD}(x_p,\xi_p)$, 
is in LO QCD equal to the ratio of the 
SD to ND structure functions of the proton. 
\begin{figure}[t]
\vglue -0.7cm
\centerline{\psfig{figure=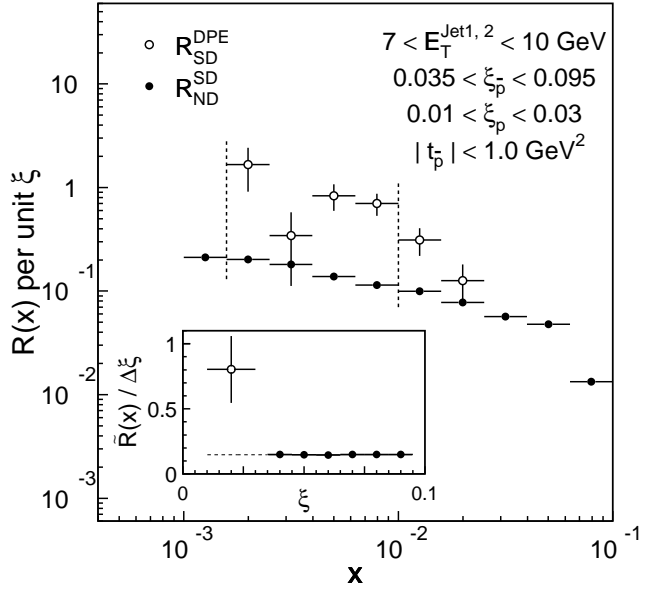,width=3.5in}}
\vglue -1cm
\caption{Ratios of DPE to SD (SD to ND) dijet event rates per
unit $\xi_p$ ($\xi_{\bar{p}}$), shown as open (filled) circles,
as a function $x$-Bjorken of partons
in the $p$ ($\bar{p}$).
The errors are statistical only. The SD/ND ratio has a normalization
systematic uncertainty of $\pm 20\%$.
The insert shows $\tilde{R}(x)$ per unit $\xi$ versus $\xi$,
where the tilde over the $R$ indicates the weighted average of
the $R(x)$ points in the region of $x$ within the vertical dashed lines,
which mark the DPE
kinematic boundary (left) and the value of
$x=\xi_p^{min}$ (right).}
\label{fig:dpe4}
\vglue -0.7cm
\end{figure}
Therefore, diffractive factorization can be tested by comparing this ratio with 
the SD to ND ratio, $R^{SD}_{ND}(x_p,\xi_p)$, for SD events  
with no rapidity gap on the antiproton side.
Since no such events are available, the comparison was made with 
the measured~\cite{CDF_JJ} ratio $R^{SD}_{ND}(x_p,\xi_{\bar p})$.
The result is shown in Fig.~\ref{fig:dpe4}.
The vertical dashed lines mark the DPE
kinematic boundary (left) and the value of
$x=\xi_p^{min}$ (right).
The weighted average of the DPE/SD points in the region
within the vertical dashed lines is $\tilde{R}^{DPE}_{SD}=0.80\pm 0.26$.
As mentioned above, 
factorization demands that $\tilde{R}^{DPE}_{SD}$ be the same
as $\tilde{R}^{SD}_{ND}$ at fixed $x$ and $\xi$.
Since the $\xi_p$ and $\xi_{\bar p}$ regions, which are respectively relevant
for the DPE/SD and SD/ND ratios, do not overlap,
we examine in the inset of Fig.~\ref{fig:dpe4}
the $\xi$ dependence of the ratios $\tilde{R}(x)$ (per unit $\xi$),
where the tilde over the $R$  indicates
the weighted average of the points in the region of $x$
within the vertical dashed lines in the main figure.
The ratio $\tilde{R}^{SD}_{ND}$, shown in six $\xi$ bins in the region
$0.035<\xi<0.095$, is flat in $\xi$. A straight line fit to the
six $\tilde{R}^{SD}_{ND}$ ratios extrapolated to $\xi=0.02$
yields $\tilde{R}^{SD}_{ND}=0.15\pm 0.02$.
The ratio of $\tilde{R}^{SD}_{ND}$ to $\tilde{R}^{DPE}_{SD}$
is $D\equiv \tilde{R}^{SD}_{ND}/\tilde{R}^{DPE}_{SD}=0.19\pm 0.07$.
The deviation of $D$ from unity represents a breakdown of factorization.

In Fig.~\ref{fig:dpe1}, the presence of the rapidity gap on the antiproton side 
reduces the rapidity range over which a gap can be formed on the proton side.
Thus, it appears that $D$ decreases
as the $\eta$-range available for the
formation of a rapidity gap increases. 
This behaviour was quantitatively predicted by 
the (re)normalized gap probability model~\cite{R}.

\section{CONCLUSIONS}
The central issue in hard diffractive production is the question of the
existence of a unique, process independent diffractive structure function (DSF).
This question has been addressed by CDF in three types of studies:

(a) Using forward rapidity gap (RG) tagging, the SD to ND ratios, $R^{SD}_{ND}$,
for $W$, dijet, $b$ and $J/\psi$ production were found to be approximately the  
same in all cases, indicating that the partonic composition of the Pomeron 
(quark to gluon fraction) is similar to that of the proton. The measured 
values of $R^{SD}_{ND}$ at $\sqrt s=1800$ GeV are $\sim 10$ times smaller 
compared to predictions based on HERA measurements, indicating a 
breakdown of factorization.

(b) Using leading antiproton (LA) tagging, the DSF was measured from 
dijet production as a function of $\xi$ and $x_{bj}$. The results 
disagree both in shape and normalization with predictions based on 
extrapolations from HERA measurements, confirming the breakdown of 
factorization found in the RG studies and extending it to the shape of the DSF.

(c) Using a combination of LA tagging [the same data sample as in (b)] 
and RG tagging on the proton side, a DPE signal was observed and 
the DSF was measured on the proton side.
Comparing the proton DSF in DPE with the $\bar p$ DSF measured in SD,
a breakdown of factorization was observed, similar in magnitude to that 
observed in SD between Tevatron and HERA.

In pursuing the underlying physics reason for the factorization breakdown in 
diffractive processes, it is interesting to note that the suppression
in the normalization of the DSF increases with the rapidity phase space 
available for RG formation, 
as foreseen in the RG renormalization model~\cite{R}.   


\begin{thebibliography}{9}
\bibitem{credits} This article contains excerpts from a paper presented by this 
author at the  ``International Symposium of Multiparticle Dynamics, Frascati,
Italy, 8-12 September 1997" [Nuc. Phys. B (Proc. Supp) 71 (1999) 368-377] 
and from Refs.~\cite{CDF_JJ,CDF_DPE}.
\bibitem{Regge} See P.D.B. Collins, {\em An Introduction to Regge Theory and 
High Energy Physics} (Cambridge University  Press, Cambridge 1977).
\bibitem{CDF_W}F. Abe {\em et al.}, CDF Collaboration,
{\em Observation of diffractive W-boson production at the Fermilab Tevatron}, 
Phys. Rev. Lett. {\bf 78} (1997) 2698.
\bibitem{CDF_JJG}F. Abe {\em et al.}, CDF Collaboration,
{\em Measurement of diffractive dijet production
at the Fermilab Tevatron}, 
Phys. Rev. Lett. {\bf 79} (1997) 2636.
\bibitem{CDF_B} T. Affolder {\it et al.}, CDF Collaboration, 
{\em Observation of diffractive $b$-quark production at the Fermilab Tevatron}, 
Phys. Rev. Lett. {\bf 84} (2000) 232.
\bibitem{CDF_JJ}T.~Affolder {\it et al.}, CDF Collaboration,
{\em Diffractive dijets with a leading antiproton in $\bar pp$ collisions 
at $\sqrt s=1800$ GeV},
Phys. Rev. Lett. {\bf 84} (2000) 5043.
\bibitem{CDF_DPE}T.~Affolder {\it et al.}, CDF Collaboration,
{\em Dijet production by double Pomeron exchange at the 
Fermilab Tevatron}, Phys. Rev. Lett. (accepted for publication). 
\bibitem{H1} H1 Collaboration, T.~Ahmed {\em et al.},
Phys. Lett. {\bf B 348} (1995) 681; C.~Adloff {\em et al.},
Z. Phys. {\bf C 76} (1997) 613.
\bibitem{ZEUS} ZEUS Collaboration,
M. Derrick {\em et al.}, Z. Phys. {\bf C 68}, (1995) 569;
Phys. Lett. {\bf B 356} (1995) 129; Eur. Phys. J. {\bf C 6} (1999) 43.
\bibitem{CDF}F. Abe {\em et al.}, CDF Collaboration, {\em The Collider Detector 
at Fermilab}, 
Nucl. Instrum. Meth. {\bf A 271} (1988) 387;
Amidei {\em et al.}, ib. {\bf A 350} (1994) 73.
\bibitem{POMPYT} P. Bruni and G. Ingelman,
in {\em Proceedings of the International Europhysics Conference
on High Energy Physics, Marseille, France, 22-28 July 1993}, 
edited by J. Carr and M. Perrottet (Editions Fronti\`{e}res, 
Gif-sur-Yvette, France, 1994) p.595.
\bibitem{CDFEDT}F. Abe {\em et al.}, Phys. Rev. {\bf D 50} (1994) 5518;
{\bf D 50} (1994) 5535; {\bf D 50} (1994) 5550.
\bibitem{R} K. Goulianos, 
{\em Renormalization of hadronic diffraction and the structure 
of the pomeron}. Phys. Lett. {\bf B 358} (1995) 379.
%{\bf B363} (1995) 268.
\end{thebibliography}
\end{document}